 \documentclass[prl,aps,twocolumn,showpacs]{revtex4}
\usepackage[dvips]{graphicx}
\usepackage{dcolumn}
\newcommand{\be}{\begin{equation}}
\newcommand{\ee}{\end{equation}}
\newcommand{\bea}{\begin{eqnarray}}
\newcommand{\eea}{\end{eqnarray}}
\newcommand{\ba}{\begin{array}}
\newcommand{\ea}{\end{array}}
\newcommand{\nn}{\nonumber}
\newcommand{\ds}{\displaystyle}
\begin{document}

\title{On the low frequency electromagnetic waves in quasi-two-dimensional metals }

\author{ Natalya A. Zimbovskaya$^{1}$ and  Godfrey Gumbs$^2$}
\affiliation{
$^1$Department of Physics and Electronics, University of Puerto Rico, CUH Station, Humacao,  PR 00791; \\
$^2$Department of Physics and Astronomy
Hunter College of CUNY, 
695 Park Avenue, New York, NY 10021  }

\begin{abstract}
Here, we theoretically analyze spectra of weakly attenuated electromagnetic waves which may appear in a Fermi-liquid of charge carriers in quasi-two-dimensional (Q2D)       layered conductors when an external magnetic field is applied perpendicularly to the conducting layers. We study transverse modes propagating along the magnetic field. The frequencies of the modes are assumed to be lower than the cyclotron frequency of the charge carriers. It is shown that Fermi-liquid interaction of the charge carriers in Q2D conductors gives rise to a mode which cannot appear in a gase of charged quasiparticles, as well as it happens in conventional metals. Also, we show that the Fermi surface plofile may cause significant changes in the waves spectra and we analyze these changes.
\end{abstract}

\date{\today}
\maketitle

In the last two decades various quasi-two-dimensional (Q2D) materials with metallic type conductivity such as organic metals, intercalated compounds and some other, have been synthesized. Galvanomagnetic phenomena and quantum oscillatory effects in these materials were intensely studied (see e.g. Refs. \cite{1,2,3,4,5}). It is a common knowledge that manifold weakly attenuating electromagnetic waves may appear in the electron liquid of conventional three dimensional metals in the presence of a strong magnetic field \cite{6,7}.
The conditions for the occurence of the electromagnetic waves may be much more favorable in Q2D conductors than in conventional metals. This happens due to a strong anisotropy of electrical conductivity in such materials. As recently shown \cite{5}, the specifics of the Q2D electron spectrum may cause peculiar features in the dispersion of some collective modes whose frequencies $ \omega $ are lower or of the order of the cyclotron frequency of the charge carriers $ \Omega. $

In the present work we mostly concentrate on the effect of the Q2D electron spectrum on the transverse Fermi-liquid (FL) cyclotron mode first predicted by Silin \cite{8}. The very existence of this mode results from the FL correlations of conduction electrons. In a metal with a nearly spherical Fermi surface (FS) this wave is circularly polarized, and it propagates along the applied magnetic field. We start our consideration by using the simple tight-binding approximation for the Q2D electron spectrum:
   \be 
 E({\bf p}) = \frac{{\bf p}_\perp^2}{2m} - 2\zeta \cos \left(p_z \frac{d}{\hbar}\right).
  \ee
   Here, $z$ axis is assumed to be perpendicular to the conducting layers; $ {\bf p}_\perp,\ p_z $ are the electron quasimomentum projections on the layer plane and $ z$-axis, respectively; the effective mass $ m $ corresponds to the electron motions in the layer plane; $ d $ is the interlayer distance. The parameter $ \zeta $ in the Eq. (1) is the interlayer transfer integral. We restrict our analysis with strong but nonquantizing magnetic fields $ \hbar/\tau \ll \hbar \Omega \ll T \ (\tau $ is the electron scattering time, and $ T $ is the temperature expressed in energy units).

Within the phenomenological Fermi-liquid theory electron-electron interactions are represented by a self-consistent field affecting any single electron included in the electron liquid. Due to this field the electron energies $ E\bf (p) $ get renormalized, and the renormalization corrections depend on the electron position $ \bf r $ and  time $ t: $
   \be 
 \Delta E = T r_{\sigma'} \int \frac{d^3 \bf p'}{(2 \pi \hbar)^3}\, F ({\bf p, \hat \sigma; p', \hat \sigma') \delta \rho (p', r',\hat\sigma',} t) .
  \ee
  Here, $ \delta \rho {\bf (p, r, \hat\sigma,} t) $ is the electron density matrix, $ \bf p $ is the electron quasimomentum, and $ \hat\sigma $ are spin Pauli matrices. The trace is taken over spin numbers $\sigma. $ The Fermi-liquid kernel included in Eq. (2) is known to have a form:
   \be 
 F ({\bf p, \hat\sigma; p', \hat\sigma') = \varphi (p,p')} + 4 \bf (\hat\sigma \hat\sigma') \psi (p,p')
  \ee 
 For an axially symmetric FS the functions $ \bf \varphi (p,p') $ and $ \bf \psi (p,p') $ actually depend on the angle $ \theta $ between $ \bf p_\perp $ and $ \bf p_\perp '$ and on the longitudinal components $ p_z, p_z'. $ Accordingly,
we may present $ \bf \varphi (p,p') $ as follows \cite{9}:
    \be
\varphi (p_z,p_z',\cos \theta) = \varphi_{00} + p_z p_z'\varphi_{01} + ({\bf p_\perp p_\perp'}) (\varphi_{10} + p_z p_z'\varphi_{11}). 
  \ee
  The function $ \psi \bf (p,p')$ may be presented in the similar way. In the Eq. (4) the functions $ \varphi_{00},\varphi_{01}, \varphi_{10},\varphi_{11} $ are even in all their arguments, namely: $ p_z,p_z'$ and $ \cos \theta .$ 

Also, we assume that both magnetic field $ {\bf B} = (0,0,B) $ and the wave vector of the considered wave $ {\bf q} = (0,0,q) $ are directed along the FS axis. Then the response of the Q2D electron liquid could be expressed in terms of the electron conductivity circular components $ \sigma_\pm (\omega, {\bf q}) = \sigma_{xx} (\omega, {\bf q}) \pm i\sigma_{yx} (\omega, {\bf q}) . $  Substituting the expression (4) for the FL function $ \bf \varphi (p,p') $ into the linearized transport equation for the nonequilibrium distribution function for the conduction electrons $ g ({\bf p,r,} t) = T r_\sigma \big(\delta \rho ({\bf p,r,\hat\sigma}, t)\big)$ one may derive the expressions for $ \sigma_\pm (\omega, \bf q) $ including the terms originating from the FL interactions. As shown in earlier works \cite{9,10}, for an axially symmetrical FS and within the adopted geometry (both $ \bf q $ and $ \bf B $ being in parallel with the FS symmetry axis) the conductivity components $ \sigma_\pm (\omega,\bf q) $ take the form:
   \be 
 \begin{array}{ll}
 \sigma_\pm& \!= \ds \frac{ 2\pi ie^2 p_F^2}{(2\pi\hbar)^3 q}\\\\
 &\!\times \frac{\ds\Phi_0^\pm \left(1-\frac{\alpha_2 u}{Q_2} \Phi_2^\pm \right)+ \frac{\alpha_2 u}{Q_2} (\Phi_1^\pm)^2}
{\ds\left[\!\left(1- \frac{\alpha_1 u}{Q_0} \Phi_0^\pm \right)\! \left( 1- \frac{\alpha_2 u}{Q_2} \Phi_2^\pm \right) \!-
\frac{\alpha_1\alpha_2}{Q_0 Q_2}(\Phi_1^\pm)^2 u^2\right]}.
  \end{array}  
  \ee
  Here,
  \bea 
 && \Phi_n^\pm = \int_{-1}^1 \frac{\overline a(x) x^n dx}{u \chi_\pm \mp \overline v (x)};
  \\ \nn \\
  &&Q_n = \int_{-1}^1 \overline a (x) x^n dx;
  \eea
  \bea 
 \chi_\pm = 1 \pm {\Omega}/{\omega} + {i}/{\omega\tau};&   \ x = {p_z}/{p_0};  &  \ u = {\omega}/{q v_0}; \nn \\ \nn\\
  \overline a (x) = {A(x)}/{A_0}; & \qquad  \overline v (x) = {v_z}/{v_0}; & 
  \eea
  where $ v_0, p_0 $ are the maximum values of the longitudinal components of the charge carriers velocity and quasimomentum, respectively: $ A_0 = \pi p_F^2;\ p = mv_F = \sqrt{2mE_F};\ E_F $ is the Fermi energy. The coefficients $ \alpha_{1,2} $ characterize FL interactions of the charge carriers:
  \bea 
 && \alpha_{1,2} = \frac{f_{1,2}}{1 + f_{1,2}};\nn\\
 && f_1 = \frac{2m}{(2\pi \hbar)^3} \int_{-p_0}^{p_0} p_\perp^2 \varphi_{10} dp_z;  \nn\\
  && f_2 = \frac{2m}{(2\pi \hbar)^3} \int_{-p_0}^{p_0} p_\perp^2 \varphi_{11} p_z^2 dp_z.
   \eea

When the charge carriers spectrum is described with the Eq. (1), the integrals $ \Phi_n^\pm, \ Q_n $ take on the form:
  \bea 
 &&\Phi_n^\pm = \int_{-1}^1 \frac{[1+ \eta \cos(\pi x)]x^n dx}{u \chi_\pm \mp \sin (\pi x)} ; \\
 && Q_n = \int_{-1}^1 [1+ \eta \cos(\pi x)]x^n dx;
  \eea
 where $ \eta = 2 \zeta/E_F \ll 1. $ Also, $ v_0 = \pi\eta v_F p_F/p_0 \ll v_F. $  

The dispersion equation for transverse modes propagating along the magnetic field could be writen as follows:
  \be 
  c^2 q^2 - 4\pi i\omega \sigma_\pm (\omega, {\bf q}) = 0.
  \ee
  When the relevant charge carriers are electrons, this equation has solutions corresponding to weakly attenuating helicoidal and Fermi-liquid cyclotron modes for $ ``-" $ polarization. Assuming the charge carriers to be holes, one must choose $``+" $ polarization to get similar solutions. For certainty, we further analyze collective modes in the electron liquid.

Using the expressions (10), (11) we may straightforwardly calculate the integrals $ \Phi_n^- $ and $ Q_n. $ For instance, we get \cite{11}:
  \be 
 \Phi_0^- = -\left((u\chi_-)^2-1\right)^{-1/2}.
  \ee
  Then, substituting the obtained results into the equation (12) we find the solutions corresponding to the low frequency waves within the collisionless limit $ (\tau \to \infty):$
  \bea 
  && \tilde\omega_1 \approx \frac{1 - \sqrt{\tilde q^2 - (\tilde q^2 - 1)(f_1 - b^2/ \tilde q^2)^2}}{1 - (f_1 - b^2/\tilde q^2)^2}; \\\nn\\
 && \tilde \omega_2 \approx (1 +f_2) \left(1 + \frac{\mu}{f_2} \tilde q^2 \right).
  \eea
  Here,
  \bea 
  \mu &=& \frac{1}{Q_2} \left [\int_{-1}^1 \! \overline a (x) \overline v^2 (x) x^2 dx - \frac{1}{Q_0} \left(\int_{-1}^1 \! \overline a (x) \overline v (x) x dx \right)^2  \right] \nn\\
  &\approx& \frac{1}{2} \left(1 - \frac{15}{2\pi^2} \right);
  \eea 
          $\ds
b^2 =\frac{\omega_p^2}{\Omega^2} \frac{v_0^2}{c^2} \frac{mv_0}{p_0};\quad \tilde q= {qv_0}/{\Omega};\quad  \tilde\omega ={\omega}/{\Omega}; $   and $ \omega_p $ is the electron plasma frequency. These  solutions describe weakly attenuated collective modes in the Fermi-liquid of a Q2D conductor provided that the FS warping is not too small $(\eta > (ql)^{-1} ,$ where $ l = v_F \tau $ is the electrons mean free path in the conduction layers planes). Within the limit $ \eta \to 0 $ which corresponds to a 2D metal with a purely cylindrical FS, the expressions (14), (15) are reduced to $ \tilde \omega_1 = 1 + f_1,\ \tilde \omega_2 = 1 + f_2, $ respectively, and they do not describe any electromagnetic waves. This is reasonable, for electromagnetic waves cannot travel in the electron liquid of a layered conductor perpendicularly to the layers when the interlayer transfer integral $ \zeta $ (see Eq. (1)) becomes zero. 

The solution (14) agrees with the corresponding result obtained by Kirichenko  et al (see Ref. \cite{5}). It describes a helicoidal wave whose spectrum at small $q\ (\tilde q \ll 1) $ has a form:
   \be 
\tilde\omega_1 = \left[\frac{b^2}{\tilde q^2} + 1 - f_1\right]^{-1} \equiv \left[\frac{\omega_p^2}{c^2q^2} \frac{mv_0}{p_0} + 1 - f_1\right]^{-1}.
  \ee
  The spectrum of the wave is changed due to the FL interactions whose effect leads to the occurence of the term $ f_1 $ in Eqs. (14), (17). In conventional 3D metals $ v_0 \sim v_F, $ therefore $ mv_0/p_0 \sim 1. $ So, assuming $ \omega_p^2/c^2q^2 \gg1, $ we may arrive at the well known expression for the frequency of the helicoidal wave: $\omega = \Omega c^2 q^2 /\omega_p^2 $ \cite{12}.
 The helicoidal wave may appear in an electron gas, and FL interactions manifest themself only through slight changes in the wave dispersion.

 On the contrary, another solution of the dispersion equation describes a mode which exists due to the FL interactions, and cannot occur in the electron gas. The edge frequency of this mode is close to the cyclotron frequency: $\omega(0) = \Omega(1 + f_2). $ As already mentioned, such Fermi-liquid cyclotron modes were predicted and their spectra analyzed by Silin for an isotropic electron liquid.  Within the isotropic model $ v_0 = v_F,\ p_0 = p_F,\ \mu = 8/35, $ so the expression (15) matchs the corresponding results of Ref. \cite{8}.
For certainty, we further assume that the FL parameter $f_2 $ takes on a negative value. Then the frequency of the Fermi-liquid cyclotron wave is smaller that the cyclotron frequency $ \Omega. $

\begin{figure}[t]
\begin{center}
\includegraphics[width=8.6cm,height=4.6cm]{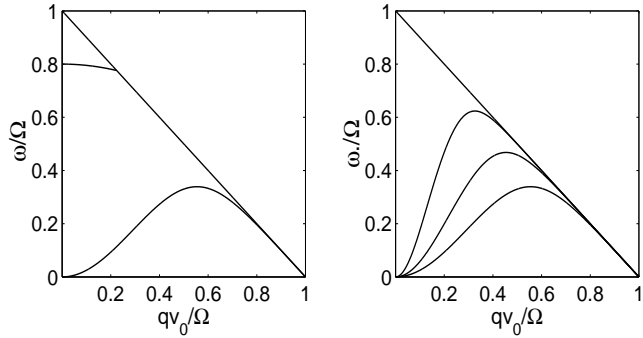}
\caption{Dispersions of low frequency transverse electromagnetic waves traveling along the magnetic field directed perpendicularly to the conducting layers in a Q2D conductor. The diagonal straight line corresponds to the Doppler-shifted cyclotron resonance. Left panel: The helicoidal (lower curve) and the Fermi-liquid (upper curve) modes. The curves are plotted using Eqs. (14), (15) for $\eta = 0.075,\ f_1 = 0.2,\ f_2 = -0.2,\ B = 1T .$ Right panel: The dispersion curves for helicoidal waves at various values of the parameter $ \eta. $ The curves are plotted for $ \eta = 0.025,\ 0.05,\ 0.075 $ (from the left to the right), at $ \alpha_1 = 0.2,\ B = 1T. $}
 \label{rateI}
\end{center}
\end{figure}

Both solutions of the dispersion equation (12) are shown in the Fig. 1 (left panel). The diagonal straight line in the figure corresponds to the Doppler-shifted cyclotron resonance $ \tilde \omega = 1 - \tilde q. $ This line separates the region where the collisionless Landau damping mechanism works (the domain above the line) from the region where it does not work (the domain below the line), and where weakly attenuating electromagnetic waves may occur. The curves in 
the figure are plotted according to Eqs. (14), (15) assuming $ f_1 =0.2,\ f_2 = - 0.2,\ \eta = 0.075 $ for $ B = 1T. $ As shown in the figure, the dispersion curve for the Fermi-liquid mode is located within the region free from the collisionless damping at small $ \tilde q. $ When $ \tilde q $ reachs the cutoff value$\tilde q_m \sim |\alpha_2| ,$ the curve meets the demarcation line $ \tilde \omega = 1 - \tilde q, $ and it terminates here. Beyond this line the wave is damped due to Landau collisionless attenuation. As for helicoidal mode, its frequency is significantly smaller than $ \Omega, $ and the wave vector may accept any value within the range $ 0 <\tilde q < 1. $ At $ \tilde q \sim 1 $ the dispersion curve of this mode is located very close to the line $ \tilde \omega = 1 - \tilde q $ which corresponds to the Doppler-shifted cyclotron resonance. Electron collisions give rise to the attenuation of the wave. However, the attenuation remains weak when $ (\Omega\tau)^{-1} <\tilde q < 1 -(\Omega\tau)^{-1}. $

\begin{figure}[t]
\begin{center}
\includegraphics[width=8.6cm,height=4.6cm]{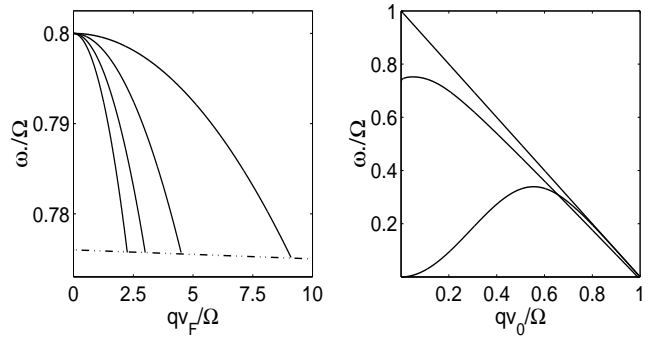}
\caption{The effect of the FS shape on the dispersion of the Fermi-liquid cyclotron waves. The curves are plotted at $ \alpha_1 = 0.2,\ \alpha_2 = -0.2,\ B = 1T. $ Left panel: Fermi-liquid cyclotron waves at various values of the parameter $ \eta $ for the cosine-like rippled FS corresponding to the energy-momentum relation (1). The curves are plotted for $ \eta = 0.1,\ 0.075,\ 0.05 $ and $ 0.025 $ (from the left to the right). The dash-dot line indicates values of the cutoff wave vector $ q_m $ for all dispersion curves. Right panel: The helicoidal (lower curve) and the Fermi-liquid (upper curve) waves in a Q2D metal with a complex FS profile. The curves are plotted using Eq. (20) with $ s= 2$ to describe $ \overline v(x), $ at $ f_1 =0.2,\ f_2 =-0.2,\ \eta = 0.075. $}
 \label{rateI}
\end{center}
\end{figure}

The effect of the FS shape on the dispersion curves of helicoidal modes is shown in the right panel of the Fig. 1. We see that the maximum frequency becomes greater, and it is reached at smaller values of $ \tilde q $ when the FS warping decreases. In the limit $ \eta \to 0 $ the curve is to be reduced to the point $\tilde q=0;\ \tilde\omega = 1 + f_1.$ Also, the FS warping has a noticeable impact on the dispersion of the Fermi-liquid wave, as demonstrated in the Fig. 2 (left panel). Here, the dispersion curves are plotted at $ f_2 = -0.2, $ and $ \eta $ takes on values from $0.025 $ to $0.1. $ The cutoff values of the wave vector magnitude $ q_m $ correspond to the points of intersection between the dispersion curves and the dash-dot line in the figure. We see that for weaker FS warping the value of $ q_m $ increases, so the wave may appear at greater wave vectors (shorter wavelengths) provided that the magnetic field is held constant. To bette estimate the effect of a Q2D electron spectrum on the Fermi-liquid cyclotron mode, we may compare the results for the cutoff value $ q_m $ presented in the Fig. 2 with those obtained for quasi-isotropic 3D metals. It is shown that for such metals $ q_m v_F/\Omega \sim |f_2| $ (see \cite{13}). Therefore, at $ f_2 = -0.2,\ q_m v_F /\Omega \sim 0.2 $ in a quasi-isotropic metal, whereas in Q2D metals with a moderately warped FS $ q_m v_F/\Omega $ may reach values significantly greater than unity at the same value of $ f_2. $

The energy-momentum relation (1) is a simplest approximation used to describe electron spectrum in Q2D conductors. Within this model the maximum value of the longitudinal velocity is reached at $ p_z = \pm p_0/2 \ (x= \pm 1/2) $ which correspond to the inflection lines on the FS. Near this lines $\overline v (x) $ may be approximates as:
  \be 
 \overline v(x) = \pm 1 \mp \frac{\pi^2}{2} \left(x \mp{1}/{2}\right)^2.
  \ee
  In realistic Q2D metals the profiles of nearly cylindrical FSs may differ from a simple cosine warping provided by the Eq. (1). In general, neglecting the anisotropy in the layers planes, we may present the energy-momentum relation in the form \cite{5}:
  \be 
  E{\bf (p)} = \frac{{\bf p}_\perp^2}{2m} - 2 \zeta \sum_{n=1}^\infty \epsilon_n \cos\left( \frac{np_zd}{\hbar} \right)
  \ee
   where $\epsilon_n $ are dimensionless coefficients. By using this expression we get opportunity to describe various FS profiles,  varying the values of $\epsilon_n. $ In particular, assuming $\epsilon_1 = 1 $ and all remaining $ \epsilon_n = 0, $ we reduce the general expression (19) to the form (1).

Assuming that the shape of the rippled FS cylinder may differ from that described by Eq. (1) we may conjecture that in some Q2D conductors the longitudinal velocity $\overline v(x) $ near the inflection lines $ x = \pm x^* $ can be presented as follows:
  \be 
  \overline v(x) = \pm 1 \mp \rho^2 (x \mp x^* )^{2s}
  \ee 
  where $ s > 1, $ and $ \rho^2 $ is the dimensionless coefficient of the order of unity. The parameter $s$ characterizes the shape of the FS segments where the longitudinal velocity $\overline v(x) $ gets maximum/minimum values. When $ s>1 $ the relative number of the electrons moving with the maximum longitudinal speed becomes significantly greater than in the case of a cosine-like FS profile (see eq. (18)). These electrons take a major part in the formation of the transverse collective modes near the Doppler-shifted cyclotron resonance. The increase in their number brings changes in the spectra of such waves. The changes are especially significant for the Fermi-liquid mode.

The effects of the FS local geometry on the spectra of the transverse Fermi-liquid waves traveling along the magnetic field in conventional 3D metals were studied in the earlier work \cite{9}. It was shown that when $ \overline v(x) $ is described by the expression of the form similar to the Eq. (20), the dispersion curve for the Fermi-liquid mode does not terminate at $ q = q_m. $ Instead, the curve has a continuation in the region of shorter wavelengths where its frequency becomes low $(\tilde\omega \ll 1). $ These conclusions remain valid for Q2D conductors. When $s$ takes on values greater than unity, the dispersion curve of the Fermi-liquid cyclotron wave runs below the demarcation line $ \tilde\omega = 1- \tilde q $ up to $ \tilde q = 1 ,$ as shown in the figure 2 (right panel). The low grequency part of the wave spectrum is located very close to this demarcation line. The latter corresponds to the Doppler-shifted cyclotron resonance, as it was mentioned above.

Finally, in the present work we studied the propagation of low frequency $(\omega <\Omega) $ electromagnetic waves in a Fermi-liquid of charge carriers in Q2D conductors. We analyzed transverse modes propagating along the magnetic field applied along the normal to the conducting layers. We found that the mode arising due to FL interactions of charge carriers may occur in Q2D metals alongside the helicoidal mode analyzed in the Ref. \cite{5}. It appears that this Fermi-liquid cyclotron mode (and the helicoidal mode, as well) may be easier observed in Q2D conductors than in usual 3D metals. This happens due to the smallness of the charge carriers velocity component along the normal to the conducting layers in Q2D metals. The smallness of $ v_z $ causes the possibility for the collective modes to exist within a broader range of the wave vector and wavelength magnitudes than in usual 3D  metals. Also, we briefly analyzed the effects of the FS local geometry in Q2D conductors on the dispersion of the Fermi-liquid cyclotron wave, and we showed that for some FS profiles the mode may exist at frequencies which are significantly lower than the cyclotron frequency.

{\it Acknowledgments:}
 We thank G. M. Zimbovsky for help with the manuscript. This work was supported  by NSF Advance program No SBE-0123654, DoD grant No W911NF-06-1-0519, and NSF-DMR-PREM 0353730.



\begin{thebibliography}{99}

\bibitem{1}  J. Wosnitza, {\it Fermi Surface of Low-Dimensional Organic Metals and Superconductors} (Springer-Verlag, Berlin, 1996).

\bibitem{2} J. Singleton, Rep. Progr. Phys. {\bf 63}, 111 (2000).

\bibitem{3} P. D. Grigoriev, Phys. Rev. B {\bf 67}, 144401 (2003).

\bibitem{4} A. Carrington, J. D. Fletcher, and T. Harima, Phys. Rev. B {\bf 71}, 174505 (2005).

\bibitem{5} O. V. Kirichenko, V. G. Peschansly, and D. I. Stepanenko, Phys. Rev. B {\bf 71}, 045304 (2005).

\bibitem{6} P. M. Platzman and P. A. Wolff, {\it Waves and Interactions in Solid State Plasma} (Academic, new York, 1973).

\bibitem{7} E. A. Kaner and V. G. Skobov, Advances in Phys. {\bf 17}, 605 (1968).

\bibitem{8} V. P. Silin, Zh. Eksp. teor. Fiz. {\bf 33}, 1227 (1958) [Sov. Phys. JETP {\bf 6}, (1958)].

\bibitem{9}  N. A. Zimbovskaya, V. I. Okulov, A. Yu. Romanov, and V. P. Silin, Fiz. Nizk. Temp. {\bf 8}, 930 (1982) [Soviet J. Low Temp. Phys. {\bf 8}, 468 (1982)].

\bibitem{10}  N. A. Zimbovskaya, Phys. Rev. B {\bf 74}, 035110 (2006).

\bibitem{11} I. S. Gradshteyn and I. M. Ryzhik, {\it Tables of Integrals, Series and Products} (Academic, New York, 2000).

\bibitem{12} A. A. Abrikosov, {\it Foundations of the Theory of Metal} (North-Holland, Amsterdam, 1988).

\bibitem{13} Y. C. Cheng, J. S. Clarke and D. Merminn, 
Phys. Rev. Lett. {\bf 20}, 1486 (1968).





\end{thebibliography}
\end{document}